\documentclass[preprint,12pt]{elsarticle}
\usepackage{graphicx}
\usepackage{amssymb}
\journal{Physics Letters A}

\begin{document}

\begin{frontmatter}

\title{Generalized algebraic transformations and exactly solvable 
classical--quantum models\tnoteref{grant}}       
\tnotetext[grant]{This work was financially supported under the grant Nos. VEGA~1/0128/08, 
VEGA~1/0431/10, and VVGS~2/09-10.}
\author{Jozef Stre\v{c}ka} 
\ead{jozef.strecka@upjs.sk}
\ead[url]{http://158.197.33.91/$\thicksim$strecka}
\address{Department of Theoretical Physics and Astrophysics, Faculty of Science,  
P. J. \v{S}af\'{a}rik University, Park Angelinum 9, 04001 Ko\v{s}ice, Slovakia}


\begin{abstract}
The rigorous approach aimed at providing exact analytical results for hybrid classical--quantum models is elaborated on the grounds of generalized algebraic mapping transformations. This conceptually simple method allows one to obtain novel interesting exact results for the hybrid classical--quantum models, which may for instance describe interacting many-particle systems composed of the classical Ising spins
and quantum Heisenberg spins, the localized Ising spins and delocalized electrons, or many other hybrid systems of a mixed classical--quantum nature.  
\end{abstract}


\end{frontmatter}

\section{Introduction}
\label{intro}

Exactly solvable models in statistical mechanics are much sought after as they often provide a valuable insight into otherwise hardly understandable aspects of interacting many-particle systems. With regard to this, exactly soluble models have become a very active research area in its own right notwithstanding the fact that rigorous solutions are usually accompanied with rather formidable 
and sophisticated mathematics \cite{domb72,thom79,baxt82,matt93,king96,lavi99,yeom02,tana02,lieb04,suth04,diep04,wu09}. 
From this point of view, it is desirable to search for simpler mathematical methods, which would 
admit to formulate and treat novel interesting exactly solvable models. 

The concept based on generalized algebraic transformations belongs to the simplest mathematical techniques, which enable a rigorous analytical treatment of diverse more complex models after performing relatively modest calculations. This rigorous method avoids formidable mathematical difficulties closely associated with an exact treatment of a more complex model by making use of  
a precise mapping equivalence with a simpler exactly solved model. An early development of 
generalized algebraic transformations has been elaborated by Fisher in the comprehensive paper 
to be published more than a half century ago \cite{fish59}. In this work, Fisher has questioned 
a possibility of how the decoration-iteration transformation, the star-triangle transformation 
and other algebraic transformations can be generalized and besides, this notable paper has also furnished a rigorous proof on a validity of algebraic mapping transformations. 

It should be pointed out that algebraic transformations are carried out at the level of partition function and their physical meaning lies in replacing a conveniently chosen part of the partition function with a simpler equivalent expression (Boltzmann's factor) to be obtained after performing a trace over degrees of freedom of in principle any decorating system. It is of fundamental importance that the trace over degrees of freedom of decorating systems can be performed independently one from each other and before summing over spin states of the Ising spins. As a result, the validity of generalized algebraic transformations can be verified locally by considering the Boltzmann's factor of the particular system, which consists of the decorating system interacting with a few outer Ising spins. Following the Fisher's argumentation, the decoration-iteration transformation for the decorating system interacting with two outer Ising spins may be of a general validity even in a presence of the external magnetic field, while the star-triangle transformation for the decorating system interacting with three outer Ising spins generally holds just in an absence of the external magnetic field provided that only pair spin-spin and single-spin interaction terms are taken into account when constructing the appropriate algebraic mapping transformation \cite{fish59}.    

Another important progress in the development of generalized algebraic transformations has recently been achieved by Rojas, Valverde and de Souza \cite{roja09} when taking into consideration a possibility of including higher-order multispin interactions in algebraic mapping transformations 
with the aim to preserve their general validity. It should be noted here that under this circumstance it is always possible to find algebraic transformation of a quite general validity 
irrespective of the total number of the outer Ising spins involved in the interactions with the decorating system. Even though the generalization invented by Rojas \textit{et al}. \cite{roja09} 
has been so far adapted only for the classical Ising models, the decorating system may be
in principle chosen in the form of arbitrary quantum-mechanical system as it has been originally 
remarked by Fisher \cite{fish59}. 

The main purpose of this Letter is to fill the gap in the theory of algebraic mapping transformations by adapting this rigorous approach for an exact treatment of classical--quantum models, which may describe hybrid systems composed of classical Ising spins and diverse decorating systems 
of a quantum nature. 

\section{Generalized algebraic transformations}

Before constructing the most general star-polygon transformation, the rigorous approach developed
by Rojas \textit{et al}. \cite{roja09} will be adapted first to treat a quite general decorating 
system interacting with either two or three outer Ising spins with the help of the generalized decoration-iteration and star-triangle transformation, respectively.

\subsection{Generalized decoration-iteration transformation}

\begin{figure}[t]
\begin{center}
\includegraphics[width=11cm]{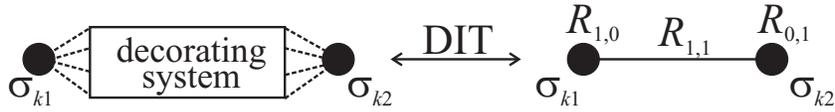}
\end{center}
\vspace{-0.6cm}
\caption{\small A diagrammatic representation of the generalized decoration-iteration transformation (DIT). By the use of this algebraic transformation, the decorating system is replaced with the effective pair interaction $R_{1,1}$ between two outer Ising spins $\sigma_{k1}$ and $\sigma_{k2}$, as well as, the generally non-uniform magnetic fields $R_{1,0}$ and $R_{0,1}$ acting on those spins.}
\label{fig:dit2}
\end{figure}

Let us consider a decorating system, which may be even of a quantum nature, coupled to the two outer Ising spins $\sigma_{k \alpha} = \pm 1/2$ ($\alpha = 1,2$) as it is schematically shown in Fig.~\ref{fig:dit2}. 
The most general Hamiltonian for any decorating system interacting with two outer  Ising spins can be formally written as 
\begin{eqnarray}
{\cal H}_k^{(2)} \left( \{ S_{ki} \}, \sigma_{k1}, \sigma_{k2} \right) 
= \!\!\! &-& \!\!\!  J_{0,0}^{(2)} \left( \{ S_{ki} \} \right) 
  -	J_{1,0}^{(2)} \left( \{ S_{ki} \} \right) \sigma_{k1}
  -	J_{0,1}^{(2)} \left( \{ S_{ki} \} \right) \sigma_{k2} \nonumber \\
  \!\!\! &-& \!\!\!	J_{1,1}^{(2)} \left( \{ S_{ki} \} \right) \sigma_{k1} \sigma_{k2}.
\label{hamdit}
\end{eqnarray}
Here, the relevant superscript marks a total number of outer Ising spins coupled with the decorating system and the symbol $\{ S_{ki} \}$ denotes all degrees of freedom of the decorating system. The physical meaning of the interaction parameters $J_{n_1, n_2}^{(2)}$ ($n_i = 0,1$ for $i = 1,2$) is quite obvious; the former (latter) subscript $n_1$ ($n_2$) determines whether or not the individual Ising spin $\sigma_{k1}$ ($\sigma_{k2}$) is indispensable part of the relevant interaction terms.

To be more specific, let us mention here few representative examples of the interaction terms represented by the interaction parameters $J_{n_1, n_2}^{(2)}$ for the particular case of the decorating system, which constitute a few quantum spins $\{ S_{ki} \}$. In this special case, 
the parameter $J_{0,0}^{(2)}$ incorporates all the interaction terms depending only on the decorating spins $\{ S_{ki} \}$. Among the most common examples of this type one could mention the interaction terms representing the effect of external magnetic field and crystal field on the decorating spins, 
as well as, bilinear, biquadratic and any other higher-order interaction terms between the decorating spins. Furthermore, the parameter $J_{1,0}^{(2)}$ involves all the interaction terms, which depend on the particular Ising spin $\sigma_{k1}$ and eventually on some of the decorating spins $\{S_{ki} \}$. This parameter may for instance represent the influence of external magnetic field on the individual Ising spin $\sigma_{k1}$, as well as, the pair and other higher-order interactions involving the Ising spin $\sigma_{k1}$ and some of the decorating spins. Similar interaction terms are also included in the parameter $J_{0,1}^{(2)}$ except that the outer Ising spin $\sigma_{k2}$ now enters into the relevant interaction terms instead of $\sigma_{k1}$. Finally, the parameter $J_{1,1}^{(2)}$ involves all the interaction terms depending on the product $\sigma_{k1} \sigma_{k2}$ between both outer Ising spins and eventually on some of the decorating spins. In this respect, the pair interaction between the two  Ising spins $\sigma_{k1}$ and $\sigma_{k2}$, as well as, the higher-order multispin interactions incorporating both these Ising spins and some of the decorating spins, are possible representatives 
of this type.

Before presenting the most general form of the decoration-iteration transformation, it is unavoidable 
to perform a trace over degrees of freedom of the decorating system in order to get the effective Boltzmann's factor $W_{k}^{(2)} (\sigma_{k1}, \sigma_{k2})$ depending merely on the two outer 
Ising spins $\sigma_{k1}$ and $\sigma_{k2}$
\begin{eqnarray}
W_{k}^{(2)} (\sigma_{k1}, \sigma_{k2}) = 
\mbox{Tr}_{\{ S_{ki} \}} \exp \left[- \beta {\cal H}_k^{(2)} 
\left( \{ S_{ki} \}, \sigma_{k1}, \sigma_{k2} \right) \right]. 
\label{rgdit1}
\end{eqnarray}
In above, $\beta = 1/(k_{\rm B} T)$, $k_{\rm B}$ is Boltzmann's constant and $T$ is the absolute temperature. It should be remarked that the former step, which bears a close relation to capability of finding the relevant trace by undertaking analytical calculation (e.g. by exact analytical diagonalization), represents perhaps the most crucial limitation to applicability of the present method to any decorating system of a quantum nature. In the latter step, the effective Boltzmann's factor (\ref{rgdit1}) can be replaced with a simpler equivalent expression depending just on the two outer Ising spins $\sigma_{k1}$ and $\sigma_{k2}$, which is provided by the generalized decoration-iteration mapping transformation
\begin{eqnarray}
W_{k}^{(2)} (\sigma_{k1}, \sigma_{k2})  
= \exp \left(\beta R_{0,0}^{(2)} + \beta R_{1,0}^{(2)} \sigma_{k1} + \beta R_{0,1}^{(2)} \sigma_{k2} 
           + \beta R_{1,1}^{(2)} \sigma_{k1} \sigma_{k2} \right). 
\label{rgdit2}
\end{eqnarray}
It is of great practical importance that the self-consistency condition of the decoration-iteration mapping transformation (\ref{rgdit2}), which ensures a general validity of this algebraic transformation irrespective of four possible spin states of the two Ising spins involved therein, allows one to express all four yet undetermined mapping parameters through the unique formula
\begin{eqnarray}
\beta R_{n_1,n_2}^{(2)} = 2^{2 (n_1 + n_2) - 2} 
\sum_{\sigma_{k1}} \sum_{\sigma_{k2}} \sigma_{k1}^{n_1} \sigma_{k2}^{n_2} 
\ln \left[ W_{k}^{(2)} (\sigma_{k1}, \sigma_{k2})\right]. 
\label{rgdit3}
\end{eqnarray}
The physical meaning of the relevant mapping parameters is as follows. The parameter $R_{1,1}^{(2)}$ 
stands for the effective pair interaction between the two outer Ising spins, the parameters $R_{1,0}^{(2)}$ and $R_{0,1}^{(2)}$ represent a generally non-uniform effective magnetic field acting 
on the Ising spins $\sigma_{k1}$ and $\sigma_{k2}$, respectively, and the last parameter 
$R_{0,0}^{(2)}$ has merely a character of the multiplicative factor. With the help of the generalized decoration-iteration transformation given by Eqs. (\ref{rgdit1}) and (\ref{rgdit2}), the Hamiltonian (\ref{hamdit}) of any decorating system interacting with the two outer Ising spins is effectively mapped onto a very simple Hamiltonian describing two mutually interacting Ising spins 
\begin{eqnarray}
\tilde{\cal H}_k^{(2)} \left( \sigma_{k1}, \sigma_{k2} \right) 
= - R_{0,0}^{(2)} - R_{1,0}^{(2)} \sigma_{k1} - R_{0,1}^{(2)} \sigma_{k2} 
  - R_{1,1}^{(2)} \sigma_{k1} \sigma_{k2}.
\label{hamdite}
\end{eqnarray}
Of course, the temperature-dependent effective interactions $R_{n_1, n_2}^{(2)}$ of the corresponding model must obey the formula (\ref{rgdit3}) stemming from the self-consistency condition of the generalized decoration-iteration transformation (\ref{rgdit1})-(\ref{rgdit2}) in order to validate the accurate mapping equivalence between both Hamiltonians (\ref{hamdit}) and (\ref{hamdite}). 

\subsection{Generalized star-triangle transformation}
\begin{figure}[t]
\begin{center}
\includegraphics[width=11cm]{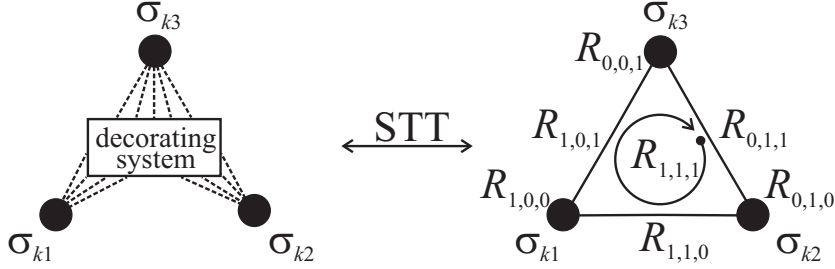}
\end{center}
\vspace{-0.6cm}
\caption{\small A diagrammatic representation of the generalized star-triangle transformation 
(STT). By the use of this algebraic transformation, the decorating system is replaced with the 
effective pair ($R_{1,1,0}$, $R_{1,0,1}$, $R_{0,1,1}$) and triplet ($R_{1,1,1}$) interactions 
between the three outer Ising spins $\sigma_{k1}$, $\sigma_{k2}$ and $\sigma_{k3}$, as well as, 
the generally non-uniform magnetic fields ($R_{1,0,0}$, $R_{0,1,0}$, $R_{0,0,1}$) acting on those spins.}
\label{fig:stt2}
\end{figure}

Even more interesting situation occurs in the search for the generalized star-triangle transformation, which could be in principle applied to any decorating system coupled to the three outer Ising spins 
$\sigma_{k \alpha} = \pm 1/2$ ($\alpha = 1,2,3$) as it is schematically illustrated in Fig.~\ref{fig:stt2}. It has been already argued by Fisher \cite{fish59} that the star-triangle transformation cannot be of a quite general validity due to a lack of the mapping parameters, which represent all possible pair spin-spin and single-spin interactions. However, the single missing mapping parameter can be chosen so as to represent the effective triplet interaction between the three outer Ising spins when including this higher-order multispin interaction into a definition of the star-triangle transformation.

The arbitrary decorating system interacting with the three outer Ising spins can be described through the most general Hamiltonian of the following form
\begin{eqnarray}
{\cal H}_k^{(3)} \!\!\!\!\!\! && \!\!\!\!\!\! 
\left( \{ S_{ki} \}, \sigma_{k1}, \sigma_{k2}, \sigma_{k3} \right) =
  - J_{0,0,0}^{(3)} \left( \{ S_{ki} \} \right) 
  -	J_{1,0,0}^{(3)} \left( \{ S_{ki} \} \right) \sigma_{k1}
  -	J_{0,1,0}^{(3)} \left( \{ S_{ki} \} \right) \sigma_{k2} \nonumber \\
  \!\!\! &-& \!\!\!	J_{0,0,1}^{(3)} \left( \{ S_{ki} \} \right) \sigma_{k3}
  - J_{1,1,0}^{(3)} \left( \{ S_{ki} \} \right) \sigma_{k1} \sigma_{k2}
  - J_{0,1,1}^{(3)} \left( \{ S_{ki} \} \right) \sigma_{k2} \sigma_{k3} \nonumber \\
  \!\!\! &-& \!\!\!	J_{1,0,1}^{(3)} \left( \{ S_{ki} \} \right) \sigma_{k3} \sigma_{k1}
  - J_{1,1,1}^{(3)} \left( \{ S_{ki} \} \right) \sigma_{k1} \sigma_{k2} \sigma_{k3}. 
\label{hamstt}
\end{eqnarray}
The physical meaning of the interaction parameters $J_{n_1, n_2, n_3}^{(3)}$ ($n_i = 0,1$ for $i = 1,2,3$) is analogous as before, namely, the notation suffix $n_i$ determines whether the individual Ising spin $\sigma_{ki}$ is present ($n_i = 1$) or absent ($n_i = 0$) in the interaction terms  represented by the relevant interaction parameter. For instance, the interaction parameter $J_{1,1,1}^{(3)}$ involves all the interaction terms depending on the product $\sigma_{k1} \sigma_{k2} \sigma_{k3}$ between all three outer Ising spins and eventually on some degrees of freedom of 
the decorating system $\{ S_{ki} \}$. 

After performing a trace over degrees of freedom of the decorating system
one obtains the effective Boltzmann's factor 
\begin{eqnarray}
W_{k}^{(3)} (\sigma_{k1}, \sigma_{k2}, \sigma_{k3}) = 
\mbox{Tr}_{\{ S_{ki} \}} \exp \left[- \beta {\cal H}_k^{(3)} 
\left( \{ S_{ki} \}, \sigma_{k1}, \sigma_{k2}, \sigma_{k3} \right) \right], 
\label{rgstt1}
\end{eqnarray}
which depends solely on the three outer Ising spins $\sigma_{k1}$, $\sigma_{k2}$ 
and $\sigma_{k3}$. The Boltzmann's factor (\ref{rgstt1}) can be subsequently replaced with 
a simpler equivalent expression, which is supplied by the generalized star-triangle mapping transformation
\begin{eqnarray}
W_{k}^{(3)} (\sigma_{k1}, \sigma_{k2}, \sigma_{k3}) \!\!\! &=& \!\!\!
 \exp \Bigl(\beta R_{0,0,0}^{(3)} + \beta R_{1,0,0}^{(3)} \sigma_{k1} 
+ \beta R_{0,1,0}^{(3)} \sigma_{k2} + \beta R_{0,0,1}^{(3)} \sigma_{k3} \nonumber \\
+ \beta R_{1,1,0}^{(3)} \sigma_{k1} \sigma_{k2} 
\!\!\! &+& \!\!\! \beta R_{0,1,1}^{(3)} \sigma_{k2} \sigma_{k3} 
+ \beta R_{1,0,1}^{(3)} \sigma_{k3} \sigma_{k1}
+ \beta R_{1,1,1}^{(3)} \sigma_{k1} \sigma_{k2} \sigma_{k3}  \Bigr). 
\label{rgstt2}
\end{eqnarray}
The generalized star-triangle transformation (\ref{rgstt2}) satisfies the self-consistency condition, which demands its general validity regardless of eight available spin configurations of the three  outer Ising spins involved therein, if and only if the mapping parameters obey the unique formula
\begin{eqnarray}
\beta R_{n_1,n_2,n_3}^{(3)} = 2^{2 (n_1 + n_2 + n_3) - 3} 
\sum_{\sigma_{k1}} \sum_{\sigma_{k2}} \sum_{\sigma_{k3}} 
\sigma_{k1}^{n_1} \sigma_{k2}^{n_2} \sigma_{k3}^{n_3}
\ln \left[ W_{k}^{(3)} (\sigma_{k1}, \sigma_{k2}, \sigma_{k3}) \right]. 
\label{rgstt3}
\end{eqnarray}
The physical meaning of the relevant mapping parameters is also quite evident. The mapping parameter  $R_{0,0,0}^{(3)}$ is an appropriate multiplicative factor, while the parameters $R_{1,0,0}^{(3)}$, $R_{0,1,0}^{(3)}$ and $R_{0,0,1}^{(3)}$ represent a generally non-uniform effective magnetic field 
acting on the Ising spins $\sigma_{k1}$, $\sigma_{k2}$ and $\sigma_{k3}$, respectively. Furthermore, the mapping parameters $R_{1,1,0}^{(3)}$, $R_{0,1,1}^{(3)}$ and $R_{1,0,1}^{(3)}$ label the effective pair interactions between three exploitable couples of the outer Ising spins and the parameter $R_{1,1,1}^{(3)}$ denotes the effective triplet interaction among them. Using the generalized star-triangle transformation 
given by Eqs. (\ref{rgstt1}) and (\ref{rgstt2}), the Hamiltonian (\ref{hamstt}) of any decorating system coupled to the three  Ising spins is effectively mapped onto the most general Hamiltonian of three mutually interacting Ising spins 
\begin{eqnarray}
\tilde{\cal H}_k^{(3)} \left( \sigma_{k1}, \sigma_{k2}, \sigma_{k3} \right) = 
\!\!\!&-&\!\!\! R_{0,0,0}^{(3)} - R_{1,0,0}^{(3)} \sigma_{k1} - R_{0,1,0}^{(3)} \sigma_{k2} 
  - R_{0,0,1}^{(3)} \sigma_{k3} - R_{1,1,0}^{(3)} \sigma_{k1} \sigma_{k2} \nonumber \\
\!\!\!&-&\!\!\! R_{0,1,1}^{(3)} \sigma_{k2} \sigma_{k3} - R_{1,0,1}^{(3)} \sigma_{k3} \sigma_{k1} 
  - R_{1,1,1}^{(3)} \sigma_{k1} \sigma_{k2} \sigma_{k3}.
\label{hamstte}
\end{eqnarray}
The rigorous mapping equivalence between both Hamiltonians (\ref{hamstt}) and (\ref{hamstte}) 
naturally demands that the temperature-dependent effective interactions $R_{n_1, n_2, n_3}^{(3)}$ 
of the corresponding model are in concordance with the formula (\ref{rgstt3}) derived from the self-consistency condition of the generalized star-triangle transformation. 

Last but not least, it should be noticed that an application of the generalized star-triangle transformation is of particular research interest just if it establishes a precise mapping equivalence with some simpler exactly solvable model. It is therefore valuable to mention few rigorously solved Ising models of this type. The spin-1/2 Ising model with only the triplet 
interaction is exactly tractable on the triangular lattice (the so-called Baxter-Wu model) \cite{wood72,merl72,baxtwu,fywu74,baxt74}, the union jack lattice \cite{hint72,fywu75,urumvi} 
and the diced lattice \cite{wood73,lliu74}. Among the more general exactly solved Ising models 
one could also mention the spin-1/2 Ising model on the diced \cite{hori85} and union jack \cite{urum88} lattices with the pair and triplet interactions, the spin-1/2 Ising model on the union jack lattice with the triplet interaction and the external magnetic field \cite{jung75,gitt80,urum86}, as well as, the spin-1/2 Ising model on the kagom\'e lattice \cite{wuwu89} and Cayley tree \cite{gani02} including 
the external magnetic field, pair as well as triplet interactions.

\subsection{Generalized star-polygon transformation}

\begin{figure}[t]
\begin{center}
\includegraphics[width=11cm]{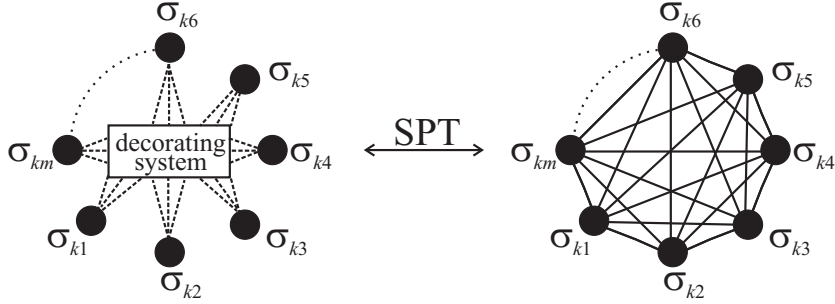}
\end{center}
\vspace{-0.7cm}
\caption{\small A diagrammatic representation of the generalized star-polygon transformation 
(SPT). Using this mapping transformation, any decorating system interacting with arbitrary
number of the outer Ising spins can be replaced with the effective interactions, which 
are represented through edges connecting each couple of the outer Ising spins, as well as, 
any polygon whose vertices all coincide with lattice points of the outer Ising spins.}
\label{fig:spt}
\end{figure}

At this stage, let us proceed to the most general star-polygon transformation, which enables to treat rigorously any decorating system interacting with arbitrary number of the outer Ising spins as it is schematically shown in Fig.~\ref{fig:spt}. It should be mentioned that the generalized star-polygon transformation involves the generalized decoration-iteration and star-triangle transformations as the special but surely the most notable cases. 
The most general Hamiltonian for any decorating system interacting with the $m$ outer Ising spins
can be written in the form
\begin{eqnarray}
{\cal H}_k^{(m)} \left( \{ S_{ki} \}, \{ \sigma_{kj} \} \right) 
= - \sum_{\{ n_j \}} J_{ \{ n_j \}}^{(m)} \left( \{ S_{ki} \} \right) 
    \prod_{j=1}^{m} \sigma_{kj}^{n_j}. 
\label{hamspt}
\end{eqnarray}
Here, the symbol $\{ n_{j} \} = \{ n_1, n_{2}, \ldots, n_{m} \}$ denotes a set of all two-valued variables, each of them determining whether the individual Ising spin $\sigma_{kj} = \pm 1/2$ is present ($n_j = 1$) or absent ($n_j = 0$) in the relevant interaction terms represented by the interaction parameter $J_{ \{ n_j \}}^{(m)}$. Next, $\{ S_{ki} \} = \{ S_{k1}, S_{k2}, \ldots, S_{kn} \}$ and $\{ \sigma_{kj} \} = \{ \sigma_{k1}, \sigma_{k2}, \ldots, \sigma_{km} \}$ denote the sets 
over all degrees of freedom of the decorating system and the outer Ising spins, respectively, and 
the summation $\displaystyle \sum_{\{ n_j \}} = \displaystyle \sum_{n_1 = 0,1} \displaystyle \sum_{n_2 = 0,1} \cdots \displaystyle \sum_{n_m =0,1}$ is carried out over the full set of two-valued variables $\{ n_{j} \}$.

The generalized star-polygon transformation can be introduced in three crucial steps. In the first step, it is necessary to perform a trace over degrees of freedom of the decorating system in order 
to get the effective Boltzmann's factor  
\begin{eqnarray}
W_{k}^{(m)} ( \{ \sigma_{kj} \}) = \mbox{Tr}_{\{ S_{ki} \}} \exp \left[- \beta {\cal H}_k^{(m)} 
\left( \{ S_{ki} \}, \{ \sigma_{kj} \} \right) \right], 
\label{rgspt1}
\end{eqnarray}
which solely depends just on the $m$ outer Ising spins from the set $\{ \sigma_{kj} \}$. 
In the second step, the Boltzmann's factor (\ref{rgspt1}) can be substituted through a simpler 
equivalent expression provided by the generalized star-polygon mapping transformation
\begin{eqnarray}
W_{k}^{(m)} ( \{ \sigma_{kj} \} ) = \exp \left[ \sum_{\{ n_j \}} \left(\beta R_{\{n_j \}}^{(m)} \prod_{j=1}^{m} \sigma_{kj}^{n_j} \right) \right]. 
\label{rgspt2}
\end{eqnarray}
In the third step, it is easy to prove that the generalized star-polygon transformation (\ref{rgspt2}) is valid independently of all spin configurations available to the outer Ising spins if and only 
if the relevant mapping parameters fulfill the unique formula
\begin{eqnarray}
\beta R_{ \{n_j \}}^{(m)} = \frac{1}{2^m} 
\sum_{ \{\sigma_{kj} \}} \Bigl[ \prod_{j=1}^m \left(4 \sigma_{kj} \right)^{n_j} \Bigr] 
\ln \left[W_{k}^{(m)} ( \{ \sigma_{kj} \} ) \right]. 
\label{rgspt3}
\end{eqnarray}
Bearing all this in mind, the generalized star-polygon transformation given by Eqs. (\ref{rgspt1}) 
and (\ref{rgspt2}) establishes a rigorous mapping equivalence between the Hamiltonian (\ref{hamspt}) 
of any decorating system interacting with the set $\{ \sigma_{kj} \}$ of the outer Ising spins 
and respectively, the most general Hamiltonian for the $m$ mutually interacting Ising spins that
includes all possible higher-order multispin interactions
\begin{eqnarray}
\tilde{\cal H}_k^{(m)} \left( \{ \sigma_{kj} \} \right) = - \sum_{\{ n_j \}}
R_{\{n_j \}}^{(m)} \prod_{j=1}^{m} \sigma_{kj}^{n_j}. 
\label{hamspte}
\end{eqnarray}
The rigorous mapping equivalence between both Hamiltonians (\ref{hamspt}) and (\ref{hamspte}) 
of course holds just if the temperature-dependent effective interactions $R_{\{n_j \}}^{(m)}$ of the corresponding model meet the unique formula (\ref{rgspt3}), which has been obtained from 
the self-consistency condition of the generalized star-polygon transformation. 

Before concluding, it is worthwhile to remark that the problem of finding a rigorous solution for any lattice-statistical model in which each decorating system interacts with a few outer Ising spins turns out to be equivalent with the problem of solving the corresponding model with temperature-dependent higher-order multispin interactions between the Ising spins. Another interesting observation is that 
the highest order of effective multispin interactions is in general equal to the total number 
of the outer Ising spins to which the relevant decorating system is coupled. In the consequence 
of that, the \textit{generalized star-square transformation} as applied to any decorating system 
interacting with four outer Ising spins will for instance afford a precise mapping relationship 
with the Ising model taking into consideration multispin interactions up to the fourth order. 
From this point of view, the lattice-statistical models in which the decorating system is coupled to four or more outer Ising spins are just barely fully exactly solvable due to 
a lack of exactly solved Ising models accounting for higher-order multispin interactions. 
There are however few notable exceptions among which one could mention the spin-1/2 Ising model 
with pair interactions on two square lattices coupled together by means of the quartic interaction, 
which is equivalent with a general eight-vertex model as first evidenced by Wu \cite{fywu71}, 
Kadanoff and Wegner \cite{kada71}. Albeit the general eight-vertex model is not exactly tractable, 
two valuable cases of the eight-vertex model has been rigorously solved under the special constraints
to its Boltzmann's weights. The symmetric eight-vertex model satisfying the zero-field condition 
has exactly been solved by Baxter \cite{zf8v71,zf8v72} and the free-fermion eight-vertex model
obeying the free-fermion condition has rigorously been solved by Fan and Wu \cite{ff8v69,ff8v70}. 

\section{Conclusions and future outlooks}

In this Letter, the rigorous approach based on the grounds of generalized algebraic transformations has been elaborated with the aim to provide a relatively simple mapping method, 
which enables an exact treatment of diverse hybrid classical--quantum models. Within the framework 
of this rigorous method, the problem of finding exact solution for any hybrid classical--quantum 
model in which any quantum-mechanical decorating system interacts with a few outer Ising spins 
turns out to be equivalent with the problem of solving the corresponding Ising model with 
the effective (temperature-dependent) multispin interactions.

Up to now, the utility of generalized algebraic transformations to provide exact results from 
a precise mapping equivalence either with the Ising model or eight-vertex model has been convincingly evidenced on two different families of exactly solvable classical--quantum models. 
The Ising-Heisenberg models, which describe interacting many-particle systems composed of the classical Ising spins and quantum Heisenberg spins, surely represent the most extended class of exactly tractable models of this kind. For instance, the generalized decoration-iteration transformation has been 
used  to obtain exact results for several trimerized and tetramerized Ising-Heisenberg linear chains \cite{stre02,stre03,stre04,stre05,lisn08}, Ising-Heisenberg sawtooth chain \cite{ohan09}, Ising-Heisenberg diamond chains \cite{jasc04,cano04,cano06,valv08,cano09,stre09}, Ising-Heisenberg chain with triangular plaquettes \cite{anto09,ohan10}, as well as, some Ising-Heisenberg models on doubly decorated planar lattices \cite{stre02a,stre02b,stre04a,stre04b,cano10}. Besides, the generalized star-triangle transformation has been employed to investigate a geometric frustration 
in the class of exactly solvable Ising-Heisenberg models on the triangulated kagom\'e lattice \cite{stre08,yao08,stre09a} and on several more complex planar lattices \cite{stre06,cano07}. 
The generalized star-square transformation, which establishes a precise mapping equivalence 
with the eight-vertex model rather than the Ising model, has been used to investigate the weak-universal critical behavior of the Ising-Heisenberg model on a square-hexagon lattice \cite{valv09} and on the planar lattice of edge-sharing octahedrons \cite{stre09c,stre09d}.

In comparison with this, the family of exactly solved hybrid models describing interacting many-particle systems composed of the localized Ising spins and mobile electrons is much less numerous. To the best of our knowledge, this kind of hybrid classical--quantum models has been
exactly solved yet merely with the help of generalized decoration-iteration transformation 
for the interacting spin-electron system on the diamond chain \cite{pere08,pere09} and 
on the doubly decorated planar lattices \cite{stre09e,stre10}. Apart from the aforementioned applications, the rigorous technique based on the generalized algebraic transformations has been also adapted to treat the hybrid classical--quantum models in which the decorating system represents a single quantum spin either interacting with the transverse magnetic field \cite{jasc99,stre03a} or the biaxial (rhombic) crystal field \cite{jasc04a,jasc05a,jasc05b}. 

To conclude, the exact mapping method based on the grounds of generalized algebraic transformations offers a variety of other opportunities to deal with in the future and it is the author's hope that many intriguing and challenging classical--quantum models will be exactly tackled using this rigorous 
technique in the near future.


\begin{thebibliography}{100}
\bibitem{domb72}
C. Domb, M.S. Green, \textit{Phase Transitions and Critical Phenomena, Vol. 1}, 
Academic Press, London, 1972.

\bibitem{thom79}
C.J. Thompson, \textit{Mathematical Statistical Mechanics}, 
Princeton University Press, New Jersey, 1979.

\bibitem{baxt82}
R.J. Baxter, \textit{Exactly Solved Models in Statistical Mechanics}, 
Academic Press, New York, 1982.

\bibitem{matt93}
D.C. Mattis, \textit{The Many-Body Problem: An Encyclopedia of Exactly Solved Models 
in One Dimension}, World Scientific, Singapore, 1993.

\bibitem{king96}
C. King, F.Y. Wu, \textit{Exactly Soluble Models in Statistical Mechanics}, 
World Scientific, Singapore, 1996. 

\bibitem{lavi99}
D.A. Lavis, G.M. Bell, \textit{Statistical Mechanics of Lattice Systems, Vol. 1}, 
Springer, Berlin-Heidelberg, 1999.

\bibitem{yeom02}
J.M. Yeomans, \textit{Statistical Mechanics of Phase Transitions}, 
Oxford University Press, Oxford, 2002. 

\bibitem{tana02}
T. Tanaka, \textit{Methods of Statistical Physics}, 
Cambridge University Press, Cambridge, 2002.

\bibitem{lieb04}
B. Nachtergaele, J.P. Solovej, J. Yngvason, \textit{Condensed Matter Physics and Exactly Soluble Models}, \textit{Selecta of E.H.~Lieb}, Springer, Berlin-Heidelberg, 2004.

\bibitem{suth04}
B. Sutherland, \textit{Beautiful Models: 70 Years of Exactly Solved Quantum}  
\textit{Many-Body Problems}, World Scientific, Singapore, 2004.

\bibitem{diep04}
H.T. Diep, H. Giacomini, \textit{Frustration - Exactly Solved Frustrated Models}  
in \textit{Frustrated Spin Systems}, Ed. H.T. Diep, World Scientific, Singapore, 2004.

\bibitem{wu09}
F.Y. Wu, \textit{Exactly Solved Models: A Journey in Statistical Mechanics}, 
World Scientific, Singapore, 2009.

\bibitem{fish59} M.E. Fisher, Phys. Rev. 113 (1959) 969. 
\bibitem{roja09} O. Rojas, J.S. Valverde, S.M. de Souza, Physica A 388 (2009) 1419.
\bibitem{wood72} D.W. Wood, H.P. Griffiths, J. Phys. C: Solid St. Phys. 5 (1972) L253.
\bibitem{merl72} D. Merlini, C. Gruber, J. Math. Phys. 13 (1972) 1814. 
\bibitem{baxtwu} R.J. Baxter, F.Y. Wu, Phys. Rev. Lett. 31 (1973) 1294.
\bibitem{fywu74} R.J. Baxter, F.Y. Wu, Aust. J. Phys. 27 (1974) 357.
\bibitem{baxt74} R.J. Baxter, Aust. J. Phys. 27 (1974) 369.
\bibitem{hint72} A. Hintermann, D. Merlini, Phys. Lett. A 41 (1972) 208.
\bibitem{fywu75} F.Y. Wu, J. Phys. C: Solid St. Phys. 8 (1975) 2262.
\bibitem{urumvi} V. Urumov, in \textit{Ordering in Two Dimensions}, 
Ed. S.K. Sinha, Elsevier, New York, 1980, pp. 361--364.
\bibitem{wood73} D.W. Wood, J. Phys. C: Solid St. Phys. 6 (1973) L135.
\bibitem{lliu74} L.L. Liu, H.E. Stanley, Phys. Rev. B 10 (1974) 2958.
\bibitem{hori85} T. Horiguchi, L.L. Gon\c{c}alves, Physica A 133 (1985) 460.
\bibitem{urum88} V. Urumov, Physica A 150 (1988) 293.
\bibitem{jung75} K. J\"ungling, G. Obermair, J. Phys. C: Solid St. Phys. 8 (1975) L31.
\bibitem{gitt80} M. Gitterman, P.C. Hemmer, J. Phys. C: Solid St. Phys. 13 (1980) L329.
\bibitem{urum86} V. Urumov, Phys. Status Solidi (B) 136 (1986) K33.
\bibitem{wuwu89} X.N. Wu, F.Y. Wu, J. Phys. A: Math. Gen. 22 (1989) L1031.
\bibitem{gani02} N.N. Ganikhodzhaev, Theor. Math. Phys. 130 (2002) 419.
\bibitem{fywu71} F.Y. Wu, Phys. Rev. B 4 (1971) 2312.
\bibitem{kada71} L.P. Kadanoff, R.J. Wegner, Phys. Rev. B 4 (1971) 3989.
\bibitem{zf8v71} R.J. Baxter, Phys. Rev. Lett. 26 (1971) 832.
\bibitem{zf8v72} R.J. Baxter, Ann. Phys. 70 (1972) 193.
\bibitem{ff8v69} C. Fan, F.Y. Wu, Phys. Rev. 179 (1969) 560.
\bibitem{ff8v70} C. Fan, F.Y. Wu, Phys. Rev. B 2 (1970) 723.

\bibitem{stre02} J. Stre\v{c}ka, M. Ja\v{s}\v{c}ur, Czech. J. Phys. 52 (2002) A37. 
\bibitem{stre03} J. Stre\v{c}ka, M. Ja\v{s}\v{c}ur, J. Phys.: Condens. Matter 15 (2003) 4519. 
\bibitem{stre04} J. Stre\v{c}ka, M. Ja\v{s}\v{c}ur, M. Hagiwara, K. Minami, 
Czech. J. Phys. 54 (2004) D583.
\bibitem{stre05} J. Stre\v{c}ka, M. Ja\v{s}\v{c}ur, M. Hagiwara, K. Minami, Y. Narumi, K. Kindo, 
Phys. Rev. B 72 (2005) 024459.
\bibitem{lisn08} B.M. Lisnii, Ukr. J. Phys. 58 (2008) 708.
\bibitem{ohan09} V. Ohanyan, Condens. Matter Phys. 12 (2009) 343.
\bibitem{jasc04} M. Ja\v{s}\v{c}ur,	J. Stre\v{c}ka, J. Magn. Magn. Mater. 272-276 (2004) 984.
\bibitem{cano04} L. \v{C}anov\'a, J. Stre\v{c}ka, M. Ja\v{s}\v{c}ur, Czech. J. Phys. 54 (2004) D579. 
\bibitem{cano06} L. \v{C}anov\'a, J. Stre\v{c}ka, M. Ja\v{s}\v{c}ur, 
J. Phys.: Condens. Matter 18 (2006) 4967.
\bibitem{valv08} J.S. Valverde, O. Rojas, S.M. de Souza, J. Phys.: Condens. Matter 20 (2008) 345208.
\bibitem{cano09} L. \v{C}anov\'a, J. Stre\v{c}ka, T. Lu\v{c}ivjansk\'y, 
Condens. Matter Phys. 12 (2009) 353.  
\bibitem{stre09} J. Stre\v{c}ka, L. \v{C}anov\'a, T. Lu\v{c}ivjansk\'y, M. Ja\v{s}\v{c}ur, 
J. Phys.: Conf. Ser. 145 (2009) 012058.
\bibitem{anto09} D. Antonosyan, S. Bellucci, V. Ohanyan, Phys. Rev. B 79 (2009) 014432.
\bibitem{ohan10} V. Ohanyan, Phys. Atom. Nucl. 73 (2010) 494.
\bibitem{stre02a} J. Stre\v{c}ka, M. Ja\v{s}\v{c}ur, Phys. Rev. B 66 (2002) 174415. 
\bibitem{stre02b} J. Stre\v{c}ka, M. Ja\v{s}\v{c}ur, Phys. Status Solidi (B) 233 (2002) R12.
\bibitem{stre04a} J. Stre\v{c}ka, M. Ja\v{s}\v{c}ur, J. Magn. Magn. Mater. 272-276 (2004) 987.
\bibitem{stre04b} M. Ja\v{s}\v{c}ur,	J. Stre\v{c}ka, Czech. J. Phys. 54 (2004) D587.
\bibitem{cano10} L. \v{C}anov\'a, J. Stre\v{c}ka, Phys. Status Solidi (B) 247 (2010) 433.
\bibitem{stre08} J. Stre\v{c}ka, L. \v{C}anov\'a, M. Ja\v{s}\v{c}ur,  M. Hagiwara, 
Phys. Rev. B 78 (2008) 024427.
\bibitem{yao08}	D.X. Yao, Y.L. Loh, E.W. Carlson, M. Ma, Phys. Rev. B 78 (2008) 024428.
\bibitem{stre09a} J. Stre\v{c}ka, L. \v{C}anov\'a, J. Phys.: Conf. Ser. 145 (2009) 012012.
\bibitem{stre06} J. Stre\v{c}ka, M. Ja\v{s}\v{c}ur, Acta Phys. Slovaca 56 (2006) 65.
\bibitem{cano07} L. \v{C}anov\'a, M. Ja\v{s}\v{c}ur, J. Stre\v{c}ka,
J. Magn. Magn. Mater. 316 (2007) e352.
\bibitem{valv09}	J.S. Valverde, O. Rojas, S.M. de Souza, Phys. Rev. E 79 (2009) 041101.
\bibitem{stre09c} J. Stre\v{c}ka, L. \v{C}anov\'a, K. Minami, Phys. Rev. E 79 (2009) 051103.
\bibitem{stre09d} J. Stre\v{c}ka, L. \v{C}anov\'a, K. Minami, AIP Conf. Proc. 1198 (2009) 156.
\bibitem{pere08} M.S.S. Pereira, F.A.B.F. de Moura, M.L. Lyra, Phys. Rev. B 77 (2008) 024402.
\bibitem{pere09} M.S.S. Pereira, F.A.B.F. de Moura, M.L. Lyra, Phys. Rev. B 79 (2009) 054427.
\bibitem{stre09e} J. Stre\v{c}ka, A. Tanaka, L. \v{C}anov\'a, T. Verkholyak, 
Phys. Rev. B 80 (2009) 174410.
\bibitem{stre10} J. Stre\v{c}ka, A. Tanaka, M. Ja\v{s}\v{c}ur, J. Phys.: Conf. Ser. 200 (2010) 022059.
\bibitem{jasc99} M. Ja\v{s}\v{c}ur,	J. Stre\v{c}ka, Phys. Lett. A 258 (1999) 47.
\bibitem{stre03a} J. Stre\v{c}ka, M. Ja\v{s}\v{c}ur, J. Magn. Magn. Mater. 260 (2003) 415.
\bibitem{jasc04a} J. Stre\v{c}ka, M. Ja\v{s}\v{c}ur, Phys. Rev. B 70 (2004) 014404.
\bibitem{jasc05a} M. Ja\v{s}\v{c}ur, J. Stre\v{c}ka, Physica A 358 (2005) 393.
\bibitem{jasc05b} M. Ja\v{s}\v{c}ur, J. Stre\v{c}ka, Condens. Matter Phys. 8 (2005) 869.

\end{thebibliography}
\end{document}